\documentclass[conference]{IEEEtran}
\IEEEoverridecommandlockouts
\usepackage{cite}
\usepackage{amsmath,amssymb,amsfonts}
\usepackage{algorithmic}
\usepackage[ruled,vlined]{algorithm2e}
\usepackage{graphicx}
\usepackage{subcaption}
\usepackage{textcomp}
\usepackage{acronym}
\usepackage[dvipsnames]{xcolor}
\usepackage{float}
\usepackage{siunitx}
\usepackage{units}
\usepackage{tikz}
\usepackage{pgf}
\usepackage{pgfplots}
\usepackage{epstopdf} %converting to PDF
\usepackage{placeins}		% floating barrier images
\graphicspath{{./img/}}

\def\BibTeX{{\rm B\kern-.05em{\sc i\kern-.025em b}\kern-.08em
    T\kern-.1667em\lower.7ex\hbox{E}\kern-.125emX}}
    
\DeclareGraphicsExtensions{.jpg,.pdf,.mps,.png}
\graphicspath{{img/} {./}} %put all figures in these dirs

\begin{document}
\title{\vspace{-0.5cm}PCM-trace: Scalable Synaptic Eligibility Traces with Resistivity Drift of Phase-Change Materials 
\vspace{-0.3cm}
\thanks{}
}
%https://www.overleaf.com/2993627371drccwmnvnynk
\acrodef{ADC}[ADC]{Analog to Digital Converter}
\acrodef{ADEXP}[AdExp-I\&F]{Adaptive-Exponential Integrate and Fire}
\acrodef{AER}[AER]{Address-Event Representation}
\acrodef{AEX}[AEX]{AER EXtension board}
\acrodef{AE}[AE]{Address-Event}
\acrodef{AFM}[AFM]{Atomic Force Microscope}
\acrodef{AGC}[AGC]{Automatic Gain Control}
\acrodef{AI}[AI]{Artificial Intelligence}
\acrodef{ALD}[ALD]{Atomic Layer Deposition}
\acrodef{AMDA}[AMDA]{AER Motherboard with D/A converters}
\acrodef{ANN}[ANN]{Artificial Neural Network}
\acrodef{API}[API]{Application Programming Interface}
\acrodef{ARM}[ARM]{Advanced RISC Machine}
\acrodef{ASIC}[ASIC]{Application Specific Integrated Circuit}
\acrodef{AdExp}[AdExp-IF]{Adaptive Exponential Integrate-and-Fire}
\acrodef{BCI}[BCI]{Brain-Computer-Interface}
\acrodef{BCM}[BCM]{Bienenstock-Cooper-Munro}
\acrodef{BD}[BD]{Bundled Data}
\acrodef{BEOL}[BEOL]{Back-end of Line}
\acrodef{BG}[BG]{Bias Generator}
\acrodef{BMI}[BMI]{Brain-Machince Interface}
\acrodef{BTB}[BTB]{band-to-band tunnelling}
\acrodef{BP}[BP]{Back-propagation}
\acrodef{BPTT}[BPTT]{Back-propagation Through Time}
\acrodef{CAD}[CAD]{Computer Aided Design}
\acrodef{CAM}[CAM]{Content Addressable Memory}
\acrodef{CAVIAR}[CAVIAR]{Convolution AER Vision Architecture for Real-Time}
\acrodef{CA}[CA]{Cortical Automaton}
\acrodef{CCN}[CCN]{Cooperative and Competitive Network}
\acrodef{CDR}[CDR]{Clock-Data Recovery}
\acrodef{CFC}[CFC]{Current to Frequency Converter}
\acrodef{CHP}[CHP]{Communicating Hardware Processes}
\acrodef{CMIM}[CMIM]{Metal-insulator-metal Capacitor}
\acrodef{CML}[CML]{Current Mode Logic}
\acrodef{CMP}[CMP]{Chemical Mechanical Polishing}
\acrodef{CMOL}[CMOL]{Hybrid CMOS nanoelectronic circuits}
\acrodef{CMOS}[CMOS]{Complementary Metal-Oxide-Semiconductor}
\acrodef{CNN}[CCN]{Convolutional Neural Network}
\acrodef{COTS}[COTS]{Commercial Off-The-Shelf}
\acrodef{CPG}[CPG]{Central Pattern Generator}
\acrodef{CPLD}[CPLD]{Complex Programmable Logic Device}
\acrodef{CPU}[CPU]{Central Processing Unit}
\acrodef{CSM}[CSM]{Cortical State Machine}
\acrodef{CSP}[CSP]{Constraint Satisfaction Problem}
\acrodef{CV}[CV]{Coefficient of Variation}
\acrodef{DAC}[DAC]{Digital to Analog Converter}
\acrodef{DAS}[DAS]{Dynamic Auditory Sensor}
\acrodef{DAVIS}[DAVIS]{Dynamic and Active Pixel Vision Sensor}
\acrodef{DBN}[DBN]{Deep Belief Network}
\acrodef{DBS}[DBS]{Deep-Brain Stimulation}
\acrodef{DFA}[DFA]{Deterministic Finite Automaton}
\acrodef{DIBL}[DIBL]{drain-induced-barrier-lowering}
\acrodef{DI}[DI]{delay insensitive}
\acrodef{DMA}[DMA]{Direct Memory Access}
\acrodef{DNF}[DNF]{Dynamic Neural Field}
\acrodef{DNN}[DNN]{Deep Neural Network}
\acrodef{DoF}[DoF]{Degrees of Freedom}
\acrodef{DPE}[DPE]{Dynamic Parameter Estimation}
\acrodef{DPI}[DPI]{Differential Pair Integrator}
\acrodef{DRRZ}[DR-RZ]{Dual-Rail Return-to-Zero}
\acrodef{DRAM}[DRAM]{Dynamic Random Access Memory}
\acrodef{DR}[DR]{Dual Rail}
\acrodef{DSP}[DSP]{Digital Signal Processor}
\acrodef{DVS}[DVS]{Dynamic Vision Sensor}
\acrodef{DYNAP}[DYNAP]{Dynamic Neuromorphic Asynchronous Processor}
\acrodef{EBL}[EBL]{Electron Beam Lithography}
\acrodef{ECoG}[ECoG]{Electrocorticography}
\acrodef{EDVAC}[EDVAC]{Electronic Discrete Variable Automatic Computer}
\acrodef{EEG}[EEG]{electroencephalography}
\acrodef{EIN}[EIN]{Excitatory-Inhibitory Network}
\acrodef{EM}[EM]{Expectation Maximization}
\acrodef{EPSC}[EPSC]{Excitatory Post-Synaptic Current}
\acrodef{EPSP}[EPSP]{Excitatory Post-Synaptic Potential}
\acrodef{ET}[ET]{Eligibility Trace}
\acrodef{EZ}[EZ]{Epileptogenic Zone}
\acrodef{FDSOI}[FDSOI]{Fully-Depleted Silicon on Insulator}
\acrodef{FEOL}[FEOL]{Front-end of Line}
\acrodef{FET}[FET]{Field-Effect Transistor}
\acrodef{FFT}[FFT]{Fast Fourier Transform}
\acrodef{FI}[F-I]{Frequency-Current}
\acrodef{FPGA}[FPGA]{Field Programmable Gate Array}
\acrodef{FR}[FR]{Fast Ripple}
\acrodef{FSA}[FSA]{Finite State Automaton}
\acrodef{FSM}[FSM]{Finite State Machine}
\acrodef{GIDL}[GIDL]{gate-induced-drain-leakage}
\acrodef{GOPS}[GOPS]{Giga-Operations per Second}
\acrodef{GPU}[GPU]{Graphical Processing Unit}
\acrodef{GUI}[GUI]{Graphical User Interface}
\acrodef{HAL}[HAL]{Hardware Abstraction Layer}
\acrodef{HFO}[HFO]{High Frequency Oscillation}
\acrodef{HH}[H\&H]{Hodgkin \& Huxley}
\acrodef{HMM}[HMM]{Hidden Markov Model}
\acrodef{HRS}[HRS]{High-Resistive State}
\acrodef{HR}[HR]{Human Readable}
\acrodef{HSE}[HSE]{Handshaking Expansion}
\acrodef{HW}[HW]{Hardware}
\acrodef{IBCI}[IBCI]{Implantable BCI}
\acrodef{ICT}[ICT]{Information and Communication Technology}
\acrodef{IC}[IC]{Integrated Circuit}
\acrodef{ICL}[ICL]{Implantable Closed Loop}
\acrodef{IEEG}[iEEG]{intracranial electroencephalography}
\acrodef{IF2DWTA}[IF2DWTA]{Integrate \& Fire 2--Dimensional WTA}
\acrodef{IFSLWTA}[IFSLWTA]{Integrate \& Fire Stop Learning WTA}
\acrodef{IF}[I\&F]{Integrate-and-Fire}
\acrodef{IMU}[IMU]{Inertial Measurement Unit}
\acrodef{INCF}[INCF]{International Neuroinformatics Coordinating Facility}
\acrodef{INI}[INI]{Institute of Neuroinformatics}
\acrodef{INRC}[Intel NRC]{Intel Neuromorphic Research Community}
\acrodef{IO}[I/O]{Input/Output}
\acrodef{IoT}[IoT]{Internet of Things}
\acrodef{IPSC}[IPSC]{Inhibitory Post-Synaptic Current}
\acrodef{IPSP}[IPSP]{Inhibitory Post-Synaptic Potential}
\acrodef{IP}[IP]{Intellectual Property}
\acrodef{ISI}[ISI]{Inter-Spike Interval}
\acrodef{IoT}[IoT]{Internet of Things}
\acrodef{JFLAP}[JFLAP]{Java - Formal Languages and Automata Package}
\acrodef{LEDR}[LEDR]{Level-Encoded Dual-Rail}
\acrodef{LFP}[LFP]{Local Field Potential}
\acrodef{LLC}[LLC]{Low Leakage Cell}
\acrodef{LNA}[LNA]{Low-Noise Amplifier}
\acrodef{LPF}[LPF]{Low Pass Filter}
\acrodef{LRS}[LRS]{Low-Resistive State}
\acrodef{LSM}[LSM]{Liquid State Machine}
\acrodef{LTD}[LTD]{Long Term Depression}
\acrodef{LTI}[LTI]{Linear Time-Invariant}
\acrodef{LTP}[LTP]{Long Term Potentiation}
\acrodef{LTU}[LTU]{Linear Threshold Unit}
\acrodef{LUT}[LUT]{Look-Up Table}
\acrodef{LVDS}[LVDS]{Low Voltage Differential Signaling}
\acrodef{MD}[MD]{Medical Device}
\acrodef{MCMC}[MCMC]{Markov-Chain Monte Carlo}
\acrodef{MEMS}[MEMS]{Micro Electro Mechanical System}
\acrodef{MFR}[MFR]{Mean Firing Rate}
\acrodef{MIM}[MIM]{Metal Insulator Metal}
\acrodef{ML}[ML]{Machine Leanring}
\acrodef{MLP}[MLP]{Multilayer Perceptron}
\acrodef{MOSCAP}[MOSCAP]{Metal Oxide Semiconductor Capacitor}
\acrodef{MOSFET}[MOSFET]{Metal Oxide Semiconductor Field-Effect Transistor}
\acrodef{MOS}[MOS]{Metal Oxide Semiconductor}
\acrodef{MRI}[MRI]{Magnetic Resonance Imaging}
\acrodef{NDFSM}[NDFSM]{Non-deterministic Finite State Machine} 
\acrodef{ND}[ND]{Noise-Driven}
\acrodef{NEF}[NEF]{Neural Engineering Framework}
\acrodef{NHML}[NHML]{Neuromorphic Hardware Mark-up Language}
\acrodef{NIL}[NIL]{Nano-Imprint Lithography}
\acrodef{NLP}[NLP]{Natural Language Processing}
\acrodef{NMDA}[NMDA]{N-Methyl-D-Aspartate}
\acrodef{NME}[NE]{Neuromorphic Engineering}
\acrodef{NN}[NN]{Neural Network}
\acrodef{NRZ}[NRZ]{Non-Return-to-Zero}
\acrodef{NSM}[NSM]{Neural State Machine}
\acrodef{OR}[OR]{Operating Room}
\acrodef{OTA}[OTA]{Operational Transconductance Amplifier}
\acrodef{PCB}[PCB]{Printed Circuit Board}
\acrodef{PCHB}[PCHB]{Pre-Charge Half-Buffer}
\acrodef{PCM}[PCM]{Phase Change Memory}
\acrodef{PD}[PD]{Parkinson Disease}
\acrodef{PE}[PE]{Phase Encoding}
\acrodef{PFA}[PFA]{Probabilistic Finite Automaton}
\acrodef{PFC}[PFC]{prefrontal cortex}
\acrodef{PFM}[PFM]{Pulse Frequency Modulation}
\acrodef{PM}[PM]{Personalized Medicine}
\acrodef{PR}[PR]{Production Rule}
\acrodef{PSC}[PSC]{Post-Synaptic Current}
\acrodef{PSP}[PSP]{Post-Synaptic Potential}
\acrodef{PSTH}[PSTH]{Peri-Stimulus Time Histogram}
\acrodef{PVD}[PVD]{Physical Vapor Deposition }
\acrodef{QDI}[QDI]{Quasi Delay Insensitive}
\acrodef{RAM}[RAM]{Random Access Memory}
\acrodef{RDF}[RDF]{random dopant fluctuation}
\acrodef{RELU}[ReLu]{Rectified Linear Unit}
\acrodef{RLS}[RLS]{Recursive Least-Squares}
\acrodef{RMSE}[RMSE]{Root Mean Squared-Error}
\acrodef{RMS}[RMS]{Root Mean Squared}
\acrodef{RNN}[RNN]{Recurrent Neural Network}
\acrodef{ROLLS}[ROLLS]{Reconfigurable On-Line Learning Spiking}
\acrodef{RRAM}[R-RAM]{Resistive Random Access Memory}
\acrodef{R}[R]{Ripples}
\acrodef{SAC}[SAC]{Selective Attention Chip}
\acrodef{SAT}[SAT]{Boolean Satisfiability Problem}
\acrodef{SCX}[SCX]{Silicon CorteX}
\acrodef{SD}[SD]{Signal-Driven}
\acrodef{SDSP}[SDSP]{Spike Driven Synaptic Plasticity}
\acrodef{SEM}[SEM]{Spike-based Expectation Maximization}
\acrodef{SLAM}[SLAM]{Simultaneous Localization and Mapping}
\acrodef{SNN}[SNN]{Spiking Neural Network}
\acrodef{SNR}[SNR]{Signal to Noise Ratio}
\acrodef{SOC}[SOC]{System-On-Chip}
\acrodef{SOI}[SOI]{Silicon on Insulator}
\acrodef{SoA}[SoA]{state-of-the-art}
\acrodef{SP}[SP]{Separation Property}
\acrodef{SRAM}[SRAM]{Static Random Access Memory}
\acrodef{STDP}[STDP]{Spike-Timing Dependent Plasticity}
\acrodef{STD}[STD]{Short-Term Depression}
\acrodef{STEM}[STEM]{Science, Technology, Engineering and Mathematics}
\acrodef{STP}[STP]{Short-Term Plasticity}
\acrodef{STT-MRAM}[STT-MRAM]{Spin-Transfer Torque Magnetic Random Access Memory}
\acrodef{STT}[STT]{Spin-Transfer Torque}
\acrodef{SW}[SW]{Software}
\acrodef{TCAM}[TCAM]{Ternary Content-Addressable Memory}
\acrodef{TFT}[TFT]{Thin Film Transistor}
\acrodef{TPU}[TPU]{Tensor Processing Unit}
\acrodef{TRL}[TRL]{Technology Readiness Level}
\acrodef{USB}[USB]{Universal Serial Bus}
\acrodef{VHDL}[VHDL]{VHSIC Hardware Description Language}
\acrodef{VLSI}[VLSI]{Very Large Scale Integration}
\acrodef{VOR}[VOR]{Vestibulo-Ocular Reflex}
\acrodef{WCST}[WCST]{Wisconsin Card Sorting Test}
\acrodef{WTA}[WTA]{Winner-Take-All}
\acrodef{XML}[XML]{eXtensible Mark-up Language}
\acrodef{CTXCTL}[CTXCTL]{CortexControl}
\acrodef{divmod3}[DIVMOD3]{divisibility of a number by three}
\acrodef{hWTA}[hWTA]{hard Winner-Take-All}
\acrodef{sWTA}[sWTA]{soft Winner-Take-All}
\acrodef{APMOM}[APMOM]{Alternate Polarity Metal On Metal}
 \author{\IEEEauthorblockN{Yi\u{g}it Demira\u{g}\textsuperscript{1*},
 Filippo Moro\textsuperscript{2}, 
 Thomas Dalgaty\textsuperscript{3},
 Gabriele Navarro\textsuperscript{2},
 Charlotte Frenkel\textsuperscript{1}, 
 Giacomo Indiveri\textsuperscript{1},\\
 Elisa Vianello\textsuperscript{2}, and
 Melika Payvand\textsuperscript{1}}
 \IEEEauthorblockA{\textsuperscript{1}\textit{\ Institute of Neuroinformatics, University of Zurich and ETH Zurich, Switzerland} \\
 \textsuperscript{2}\textit{\ CEA-Leti, Grenoble, France},
 \textsuperscript{3}\textit{\ CEA-List, Nano-INNOV, Université Paris-Saclay, France} \\
 \vspace{-1.3cm}}

 }
\maketitle
\begin{abstract}
  Dedicated hardware implementations of spiking neural networks that combine the
  advantages of mixed-signal neuromorphic circuits with those of emerging memory
  technologies have the potential of enabling ultra-low power pervasive sensory
  processing. 
  To endow these systems with additional flexibility and the ability to learn to solve specific
  tasks, it is important to develop appropriate on-chip learning mechanisms.
  Recently, a new class of \emph{three-factor} spike-based learning rules have been proposed that 
  can solve the temporal credit assignment
  problem and approximate the error back-propagation algorithm on complex tasks.
  However, the efficient implementation of these rules on hybrid CMOS/memristive
  architectures is still an open challenge.
  Here we present a new neuromorphic building block, called PCM-trace, which
  exploits the drift behavior of phase-change materials to implement long lasting eligibility traces, a critical ingredient of three-factor learning rules.  
  We demonstrate how the proposed approach improves the area efficiency by $> 10\times$ compared to existing solutions and demonstrates a technologically plausible
  learning algorithm supported by experimental data from device measurements.
\end{abstract}

\vspace*{-1mm}
\section{Introduction}
%%%%% OLD VERSION %%%%%
Neuromorphic engineering uses electronic analog circuit elements to implement compact and energy-efficient intelligent cognitive systems~\cite{Mead20,Chicca_etal14,Indiveri_Horiuchi11,Mead90}.
Leveraging substrate's physics to emulate biophysical dynamics is a strong incentive toward ultra-low power and real-time implementations of neural networks using mixed-signal memristive event-based neuromorphic circuits~\cite{Chicca_Indiveri20,Serb_etal16,Li_etal18c,Spiga_etal20}.
The majority of these systems are currently deployed in edge-computing applications only in \emph{inference mode}, in which the network parameters are fixed.
However, learning in edge computing can have many advantages, as it enables adaptation to changing input statistics, reduced network congestion, and increased privacy. 
%%%%%%%%%%%%%%%%%%%%
%%% NEW VERSION %%%%%%%%%%
%To build intelligent agents continuously interacting with a changing environment, a compact, low power, adaptive, and real-time computing substrate is required. These requirements are best implemented by utilizing the substrate's physics, which naturally computes and learns based on the data it receives. 
%Thus, mixed-signal event-based neuromorphic circuits augmented by the emergence of memristive devices provide a promising solution~\cite{Payvand_etal19,Chicca_Indiveri20,Li_etal18c,Spiga_etal20}. %The majority of these systems are currently deployed in edge-computing applications only in the \emph{inference mode}, in which the network parameters are fixed.
%However, learning on the edge of these systems enables adaptation to changing input statistics along with reduced network congestion and increased privacy.
%%%%%%%%%%%%%%%%%%%%%%%%
Indeed, there have been multiple efforts implementing \ac{STDP}-variants and Hebbian learning using neuromorphic processors~\cite{Qiao_etal15,Frenkel_etal19,Payvand_Indiveri19}. These methods control \ac{LTD} or \ac{LTP} by specific local features of pre- and post-synaptic activities.
However, local learning rules themselves do not provide any guarantee that network performance will improve in multi-layer or recurrent networks.
Local error-driven approaches, e.g., the Delta Rule, aim to solve this problem but fail to assign credit for neurons that are multiple synapses away from the network output~\cite{Widrow_Hoff60, Payvand_etal20b}.
On the other hand, it has been recently shown that by using external \emph{third-factor} neuromodulatory signals (e.g., reward or prediction error in reinforcement learning, teaching signal in supervised learning), this can be achieved in hierarchical networks~\cite{Gerstner_etal18, Neftci18}. 
However, there needs to be a mechanism for synapses to remember their past activities for long periods of time, until the reward event or teacher signal is presented.
In the brain, these signals are believed to be implemented by calcium ions, or CAMKII enzymes in the synaptic spine~\cite{Sanhueza_Lisman13} and are called eligibility traces.
In machine learning, algorithmic top-down analysis of the gradient descent demonstrated how local eligibility traces at synapses allow networks to reach performances comparable to error back-propagation algorithm on complex tasks~\cite{Bellec_etal20, Zenke_Ganguli18,Rumelhart_etal86a}. 
Examples of neuromorphic platforms that implement these types of eligibility traces in spiking neural networks already exist~\cite{Davies_etal18,Furber_etal14,Grubl_etal20}.
However, learning in these platforms is only supported through the use of von-Neumann processors, either shared with the computation of network dynamics~\cite{Furber_etal14} or a dedicated core~\cite{Davies_etal18,Grubl_etal20}. Relying on numerical integration, these platforms do not leverage the physics of their computing substrate and are not free from the von-Neumann bottleneck problem~\cite{Backus78,Indiveri_Liu15}. 
On the other hand, mixed-signal analog/digital neuromorphic circuits allow the use of in-memory computing that directly emulates the desired neural and synaptic dynamics using the physics of analog elements~\cite{Qiao_etal17,Bartolozzi_Indiveri07a,Payvand_etal20a}.
However, even though progress has been made in extending the duration of synaptic traces using advanced \ac{FDSOI} technologies~\cite{Rubino_etal19}, implementing tens-of-seconds-long time constants solely based on \ac{CMOS} is not scalable, as it requires the use of large capacitors.
In this paper, we present a novel approach to exploit the drift
behavior of \ac{PCM} devices to intrinsically perform \ac{ET} computation over behavioral timescales. We present the \emph{PCM-trace} building block as a hybrid memristive-CMOS circuit solution that can lead to record-low area requirements per synapse.
To the best of our knowledge, this is the first work that uses a memristive device not only to store the weight of synapses, but also to keep track of synaptic eligibility to interact with a third factor toward scalable next-generation on-chip learning. 

\vspace{-0.2cm}
\section{Eligibility Traces}

\begin{figure}[t]
    \vspace*{-0.4cm}
	\centering
    \includegraphics[width=0.42\textwidth]{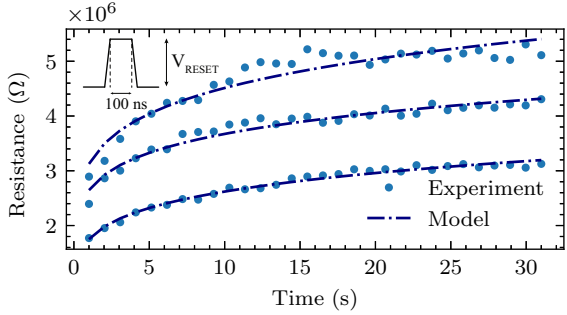}
    \caption{Experimental (dots), and simulated (dashed lines) resistance drift characteristics at constant room temperature.\vspace{-0.3cm}} %Each device is programmed to have different initial resistance values to verify Eq.~\ref{eq:drift}. \SI{0.05}{\volt} READ pulses are applied to measure low-field resistance values to not disturb the phase change material state.}
    \label{fig:data_PCM_drift}
    \FloatBarrier
    \vspace{-0.2cm}
\end{figure}
The \ac{ET} can be described as a decaying synaptic variable as in Eq.~(\ref{eq:e-trace}). 
The value of the \ac{ET} at the synapse between pre-synaptic neuron $j$ and post-synaptic neuron $i$ can be controlled as a usually linear function, $f_j$, of the pre-synaptic activity $x_j$, and potentially non-linear function $g_i$ of the post-synaptic activity $x_i$ such that

\vspace*{-5mm}
\begin{equation}
  e_{ij}^{t+\Delta t} = \alpha e_{ij}^{t} + \eta f_{j}(x_{j}^t) g_{i}(x_{i}^t),
\label{eq:e-trace}
\end{equation}

\noindent where $\eta$ is a constant and $\alpha = e^{-\Delta t / \tau_{m}}$ is
the decay rate of \ac{ET}, $\tau_{m}$ is decay time-constant up to tens of
seconds in behavioural time-scales and $\Delta t$ is discrete time-step~\cite{Gerstner_etal18}.
The $e_{ij}$ acts as a temporal correlation detector between pre-synaptic $f_j(x_j)$ and post-synaptic $g_i(x_i)$ functions. 
The instantaneous correlation between $f_j$ and $g_i$ is defined as \emph{synaptic tagging}, which is accumulated by $e_{ij}$ to keep track of past correlations. 
The $f$ and $g$ functions are determined by the chosen synaptic learning rule.
For example, $f_j(x_j)$ is the low-pass filtered pre-synaptic events in e-prop and BDSP~\cite{Bellec_etal20,Payeur.2020}, and $g_i(x_i)$ is a non-linear function of the
post-synaptic state for e-prop and SuperSpike~\cite{Zenke_Ganguli18}.

\section{PCM Measurements}
Temporal evolution of electrical resistivity is a widely-observed phenomenon in \ac{PCM} due to the rearrangements of atoms in the amorphous phase~\cite{Ielmini_etal08}. 
This behavior is commonly referred to as structural relaxation or drift.
To start the drift, a strong RESET pulse is applied to induce a crystalline to amorphous phase transition where the \ac{PCM} is melted and quenched. 
The low-ordered and highly-stressed amorphous state then evolves to a more energetically favorable glass state within tens of seconds~\cite{Le-Gallo_etal18}.
%All other conditions equal, at \ac{HRS}, the drift is stronger than \ac{LRS}.

% The model
At constant ambient temperature, the resistivity follows

\begin{equation}
	R(t)=R(t_0)\left(\frac{t}{t_0}\right)^\nu, 
	\label{eq:drift}
\end{equation}
 
\noindent where $R(t_0)$ is the resistance measured at time $t_0$ and $\nu$ is
the drift coefficient.
It has been experimentally verified by many groups that Eq.~(\ref{eq:drift}) can successfully capture
the drift dynamics~\cite{Karpov_etal07,
  Le-Gallo_etal18, Pirovano_etal04}, from microseconds to hours range~\cite{Kim_etal11a}.
 
% EDITED CEA-LETI contribution (Removed CEA-LETI due to double-blind review requirements)
We integrated Ge\textsubscript{2}Sb\textsubscript{2}Te\textsubscript{5}-based \ac{PCM} in state-of-the-art \ac{PCM} heater-based devices fabricated in the Back-End-Of-Line (BEOL) based on \SI{130}{\nano\meter} \ac{CMOS} technology. 
The \ac{PCM} thickness is \SI{50}{\nano\meter} with the bottom size of \SI{60x60}{\nano\meter\square}.
Drift measurements were performed on three devices to monitor the temporal evolution of the resistance in the \ac{HRS} state, particularly confirming the model in Eq.~(\ref{eq:drift}).
The test was conducted by first resetting all the cells by applying a RESET pulse to the heater, which has a width of \SI{100}{\nano\second} with \SI{5}{\nano\second} rising and falling times, and a peak voltage of \SI{1.85}{\volt}.
Then, an additional programming pulse is used to bring the devices to different initial conditions, corresponding to $R(t=\SI{1}{\second})=[\SI{1.77}{\mega\ohm}, \SI{2.39}{\mega\ohm}, \SI{2.89}{\mega\ohm}]$.
The low-field device resistances are measured every \SI{1}{\second} for \SI{30}{\second} by applying a READ pulse which has the same timing of the RESET pulse but a peak voltage of \SI{0.05}{\volt}.

\section{PCM-trace}
PCM-trace is a novel method to implement seconds-long \ac{ET} for the synapse using the drift feature of \ac{PCM}. 
By writing Eq.~(\ref{eq:drift}) as a difference equation of the
conductance, we can show that the temporal evolution of the conductance has
decay characteristics similar to Eq.~(\ref{eq:e-trace}) such that
$G_{ij}^{t+\Delta t}=(\frac{t-t_p}{t-t_p+\Delta t})^\nu G_{ij}^{t}$, where $G_{ij}^{t_0}=1/R_{ij}^{t_0}$, and $t_p$ is the last programming time as drift re-initializes with every gradual SET \cite{ demirag_multiphysics_2018, Nandakumar_etal18}.
%The main difference is that PCM-trace decay rate is not a constant and changes with time; nevertheless, it is comparable to the desired $\alpha$, as $\lim_{t \to \SI{10}{\second}} (\frac{t}{t+\Delta t})^\nu \approx \alpha = 0.9999$. 
The main difference is that the rate of change in PCM resistivity is a function
of time; nevertheless, its time constant is comparable for behavioral
time-scales as $\tau_{PCM}=-\Delta t/log((t/(t+\Delta t))^\nu)$ is on the order
of tens of seconds \cite{Lehmann_etal19}.
Therefore, the PCM-trace dynamics can emulate the \ac{ET} of the synapse as follows:  

\vspace*{-3mm}
\begin{equation}
	G_{ij}^{t+\Delta t}=\left(\frac{t-t_p}{t-t_p+\Delta t}\right)^\nu G_{ij}^{t} + \eta f_j(x_j^t) g_i(x_i^t)
	\label{eq:PCM-trace}
\end{equation}
\vspace*{-1mm}

% Accumulating trace and boundaries (Still keep general.)
\begin{figure}[t]
    \vspace*{-0.3cm}
	\centering
    \includegraphics[width=0.5\textwidth]{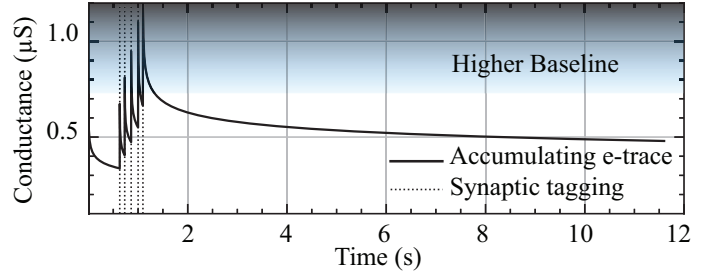}
    \caption{Accumulating \ac{ET} using PCM-trace drift
      model (Eq.~\ref{eq:PCM-trace}). After resetting the PCM-trace device at
      $t=0$, 5 random synaptic tags %(typically tag frequency is lower than pre and post firing rates) 
      are applied to the synapse, implemented by a
      gradual SET for each tag that results in $50\%$ increase in the conductivity. The device can keep the \ac{ET} for more than \SI{10}{\second}.}
    \label{fig:acc_single}
\vspace{-0.4cm}    
\end{figure}
In the PCM-trace method (Eq.~\ref{eq:PCM-trace}), the accumulating term on the \ac{ET} is implemented by applying a gradual SET to the \ac{PCM} device whenever the synapse is tagged. 
To maximize the number of accumulations a \ac{PCM} device can handle without getting stuck in the \ac{LRS} regime, some operational conditions need to be satisfied. 
We initialize the device to \ac{HRS} by applying a strong RESET pulse, and wait for an initialization time $t_{init}$ of at least $\SI{250}{\milli\second}$ for the device resistance to increase. 
If $t_{init}$ is too short, the device conductance would still be too high to be able to accumulate enough tags; and if it is too long, the decay will be weaker (see Eq.~\ref{eq:drift}).
Initialization time can be modulated to reach the desired drift speed depending on the material choice and the application.
After the initialization time, whenever the synapse is tagged, a single gradual SET (with an amplitude of \SI{100}{\micro\ampere} and a pulse width of \SI{100}{\nano\second} with \SI{5}{\nano\second} rising and falling times) is applied. 
To make sure that the device stays in the HRS, a read-verify-set scheme can be used. Finally, the value of the \ac{ET} can be measured after seconds by reading the conductance of the device (see Fig.~\ref{fig:acc_single}).

\subsection{Multi PCM-trace}
\begin{figure}
	\vspace*{-3mm}
	\centering
    \includegraphics[width=0.45\textwidth]{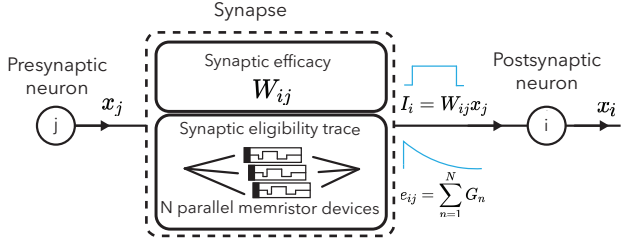}
    \caption{Multi PCM-trace concept. Each synapse has a weight and a PCM-trace block where multiple parallel \ac{PCM} devices keep the \ac{ET} of the synapse with their natural drift behavior. 
    The postsynaptic neuron receives the sum of product of the pre-synaptic activity and the weight block. 
    In parallel, the PCM-trace block calculates the \ac{ET} as a function of pre- and post-synaptic activities (Eq.~\ref{eq:PCM-trace}), to be used in the weight update.\vspace{-0.2cm}}
    \label{fig:pcm-trace-illustration}
\vspace{-0.25cm}
\end{figure}

\begin{figure}[!ht]
	\centering
    \includegraphics[width=0.5\textwidth]{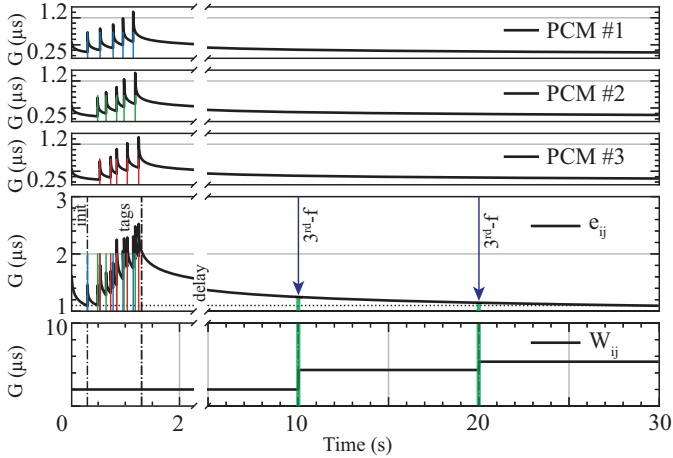}
    \caption{Accumulating \ac{ET} using multi-PCM configuration. 
    Synapse receives 15 tags between \SIrange{300}{1300}{\milli\second} which are routed to three different devices shown in the top three plots. 
    The effective \ac{ET} is calculated by applying a READ pulse to the parallel \ac{PCM} devices. 
    The initialization duration and synaptic activity period are shown with dashed lines in the bottom plot. 
    The synaptic efficacy $W_{ij}$ is modified depending on the state of \ac{ET} once the third-factor signal arrives.}
    \vspace{-0.2cm}
    \label{fig:acc_multi}
 \end{figure}

%In this section, we will demonstrate the concept of utilizing multiple PCM-trace devices to have better accumulation of the \ac{ET}. 
The number of gradual SET pulses applied to a single PCM-trace device is limited, because each pulse partially increases the device conductivity and eventually move the device toward its \ac{LRS} ($<$ \SI{2}{\mega\ohm}), where the drift converges to a higher baseline level. 
This problem can be solved by storing the synaptic \ac{ET} distributed across multiple \ac{PCM} devices, as in Fig.~\ref{fig:pcm-trace-illustration}. 
By successively routing the tags to multiple \ac{PCM} devices, the number of gradual SET pulses to be applied per single device is significantly reduced. 
Fig.~\ref{fig:acc_multi} demonstrates the increase of effective dynamic range (number of updates to \ac{ET} without getting stuck in the \ac{LRS}) using multiple \ac{PCM} devices.

\section{Circuit and Architecture}

\vspace*{0.5mm}
\subsection{PCM-trace Architecture}
\begin{figure}[t]
	\vspace*{-3mm}
	\centering
    \includegraphics[width=0.5\textwidth]{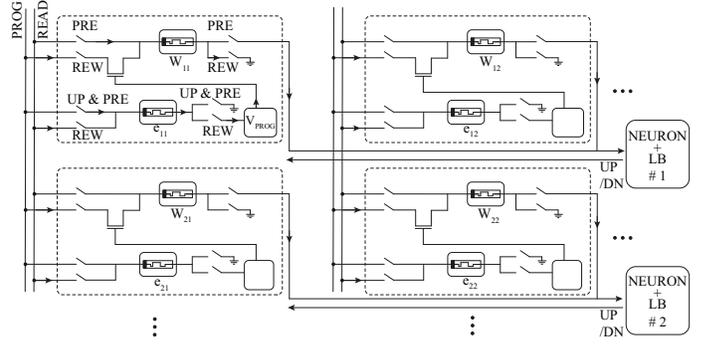}
    \caption{PCM-trace-based neuromorphic architecture for three-factor learning. Only positive \ac{ET} $(e_{ij}^+)$ and $W_{ij}^+$ are shown.}
    \label{fig:arch}
\vspace{-0.4cm}
\end{figure}

An example in-memory event-based neuromorphic architecture is shown in Fig.~\ref{fig:arch}, where the PCM-trace is employed to enable three-factor learning on behavioral time scales. 

\emph{Synapse:} Each synapse includes a weight block $W_{ij}$ in which two
%Ge\textsubscript{2}Sb\textsubscript{2}Te\textsubscript{5}-based 
PCM devices are used
in differential configuration to represent positive and negative
weights~\cite{Payvand_etal19}.
The effective synaptic weight is calculated as the difference of these two
conductance values, i.e., $W_{ij}=W_{ij}^{+} - W_{ij}^{-}$.
Also, each synapse has a PCM-trace block $e_{ij}$ that keeps the \ac{ET}.
Inside the PCM-trace block, there are two
%Ge\textsubscript{2}Sb\textsubscript{2}Te\textsubscript{5} 
PCM devices, keeping track
of the positive and negative correlation between pre and post-synaptic neurons.
On the onset of the pre-synaptic input spike, $PRE_{j}$, (i) $W_{ij}$ is read,
and the current is integrated by the post-synaptic neuron $i$; (ii) Based on the $UP$/$DN$ signal from the learning block (LB), a gradual SET programming current is applied to positive/negative
PCM-trace devices.
%(Fig.~\ref{fig:arch} only shows the positive device).

\emph{Neuron with Learning Block (LB):} The LB estimates the pre-post synaptic neuron correlation using the \ac{SDSP} rule~\cite{Brader_etal07}.
At the time of the pre-synaptic spike, the post-synaptic membrane variable is compared against a threshold, above (below) which an $UP$ ($DN$) signal is generated representing the tag type.
On the arrival of the third factor binary reward signal, $REW$, the state of the \acp{ET} devices is read by the $V_{PROG}$ block (Fig.~\ref{fig:circuits}b) which generates a gate voltage that modulates the current that programs the weight devices $W_{ij}$ (see Alg.~\ref{alg:pcm-alg}).

\vspace{-0.2cm}

\begin{algorithm}
%\SetAlgoLined
 $W_{ij}^{+} = rand(){;}$ $W_{ij}^{-} = rand()$\;
 $RESET(e_{ij}^{+}){;}$ $RESET(e_{ij}^{-})$\; 

 \While{$t < taskDuration$}{
 $I_{i,x} = 1-(V_{i,th}-V_{i,mem})/V_{i,th}$\;
  \If{@Pre \textbf{and} $t > t_{init}$}{
     $\textit{\# Eligibility trace accumulation}$\\
     \ForAll{$e_{ij}$}{
  	  \If{$I_{i,x} > I_{th}^{+}$}{
  	  	$\textit{GRADUAL\_SET}(e_{ij}^{+})$\;
  	  }
  	  \If{$I_{i,x} < I_{th}^{-}$}{
  	  	$\textit{GRADUAL\_SET}(e_{ij}^{-})$\;
  	  }
   }
  }
      $\textit{\# Third-factor}$\\
   \If{Reward}{
   \ForAll{$W_{ij}$}{
       $I_{ij,e^{+}},I_{ij,e^{-}}=READ(e_{ij}^{+}, e_{ij}^{-})$\;
       $I_{PROG}^{+}=I_{ij,e^{+}}*scale\_const$\;
       $I_{PROG}^{-}=I_{ij,e^{-}}*scale\_const$\;
  	   $\textit{GRADUAL\_SET}(W_{ij}^{+}, I_{PROG}^{+})$\;
  	   $\textit{GRADUAL\_SET}(W_{ij}^{-}, I_{PROG}^{-})$\;
   }
  }
  }
 \caption{\vspace*{-3mm}Three-factor learning with PCM-trace}
 \label{alg:pcm-alg}
\end{algorithm}

\vspace*{2mm}
\subsection{Circuit simulation}
\begin{figure}[!ht]
    \vspace*{-3mm}
 	\centering
     \includegraphics[width=0.5\textwidth]{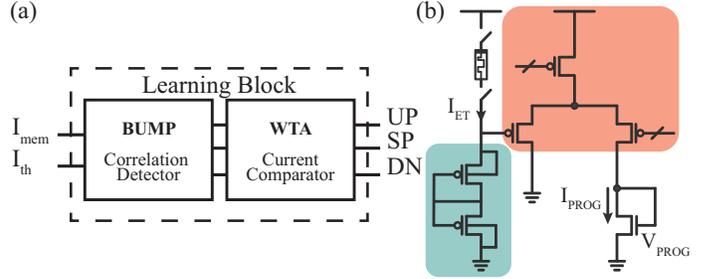}
     \caption{(a) Learning block diagram generating UP/DN signals as a function of the correlation between pre and post-synaptic activity. (b) $V_{PROG}$ circuit reading from the \ac{ET} device through the voltage divider (green) and generating $I_{PROG}$ through the diff. pair (red) to program the weight device.}
     \label{fig:circuits}
 \end{figure}

\begin{figure}[!ht]
	\centering
    \includegraphics[width=0.5\textwidth]{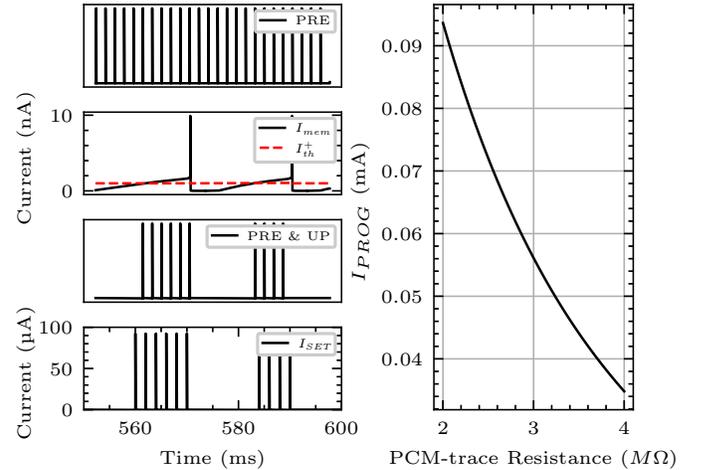}
    \caption{a) From the top: PRE events, POST membrane current ($I_{mem}$) and learning threshold ($I_{th}$), PRE events only when $I_{mem}$ is higher than $I_{th}$, and corresponding gradual SET current pulse applied to PCM-trace. b) Programming current to be applied to the weight \ac{PCM} as a function of \ac{ET} state.}
    \label{fig:circuitsim}
\vspace{-0.48cm}
\end{figure}

Fig.~\ref{fig:circuits} describes the block diagram of the LB implementing \ac{SDSP} rule, which calculates the pre-post neurons' correlation. 
The membrane variable (described here as a current $I_{mem}$ since circuits are
in current-mode) is compared against a threshold value $I_{th}$  through a Bump
circuit~\cite{Delbruck_Mead93, Payvand_etal19}.
The output of this block is digitized through a current comparator (in our
design chosen as a \ac{WTA} block~\cite{Liu_etal02a}) and generates $UP$/$DN$
signals if the membrane variable is above/below the threshold $I_{th}$, and
STOP, $SP$, if they are close within the dead zone of the bump circuit
\cite{Delbruck_Mead93}.
Fig.~\ref{fig:circuits}b presents the circuit schematic which reads the PCM-trace and generates $V_{PROG}$. 
To read the state of the device, a voltage divider is formed between the PCM device and a pseudo resistor, highlighted in green. As the device resistance changes, the input voltage to the differential pair, highlighted in red, changes. This change is amplified by the gain of the diff. pair and the device current is normalized to its tail current giving rise to $I_{PROG}$ which develops $V_{PROG}$ through the diode-connected NMOS transistor.
$V_{PROG}$ is connected to the gate of the transistor in series with the weight PCM (see Fig.~\ref{fig:arch}).\\
%We have designed these circuits in \SI{180}{\nano\meter} process and report the Spectre simulation results.  
Fig.~\ref{fig:circuitsim}a plots PRE, $I_{mem}$, the output of the learning block at the time of the PRE, and the gradual SET pulse applied to the device. 
As shown, the $UP$ signal is asserted when the membrane current is higher than the threshold indicated in red, which causes a gradual SET pulse with \SI{100}{\micro\ampere} to be applied across the PCM-trace device upon PRE events.
Fig~\ref{fig:circuitsim}b shows the generated $I_{PROG}$ as a function of the state of the \ac{ET} device. 
The higher the ET device's resistance, the less the accumulated correlation, thus the lower the programming current that should be applied to the weight device. 
The resistance on the x axis of the plot matches the measured resistance of \ac{PCM} devices shown in Fig.~\ref{fig:data_PCM_drift}.

\begin{table}[b]
\caption{Area comparison of \ac{ET} implementation}
\label{tab:et_area}
\centering
\begin{tabular}{|l|l|l|l|}

\hline

\textbf{}& \textbf{Area ($\si{\micro\meter\squared}$)} & \textbf{$\tau (\si{\second})$} & \textbf{Area/$\tau$ ($\si{\micro\meter\squared\per\second}$)} \\
\hline
\textbf{CMOS~\cite{Rubino_etal19}}  & $20 \times 17$ & $6$ & $56.6$  \\
\hline
\textbf{\ac{PCM} [This work]}  & $12\times 12$ & $>30$ & $<4.8$ \\
\hline
\end{tabular}
\end{table}


\section{Discussion and Conclusion}
% TODO: RESET the device in every 30 seconds to keep drift fast.
% Noise

Long-lasting \acp{ET} enable the construction of powerful learning mechanisms for solving complex tasks by bridging the synaptic and behavioral time-scales.
In this paper, for the first time, we proposed to use the drift of \ac{PCM} devices to implement \acp{ET}, and analyzed their feasibility for implementation in existing fabrication technologies.

%\paragraph{Scaling}
The implementation of the three-factor learning rules with \acp{ET} per synapse
requires complex memory structures for keeping track of the \ac{ET} and
the weight. Our proposed approach has clear advantages for scaling. 
Table~\ref{tab:et_area} shows a comparison between our \ac{PCM} synapse and a \ac{CMOS}-only implementation in \SI{22}{\nano\meter} \ac{FDSOI} technology from~\cite{Rubino_etal19}.

%\paragraph{Feasibility}
\ac{PCM} is among the most advanced emerging memory technology integrated into the neuromorphic domain~\cite{Boybat_etal18}.
Our approach of using \ac{PCM} to store both the synaptic weight and
the \ac{ET} requires no additional nano-fabrication methods.

\section*{Acknowledgment}
This project has received funding from the European Union's H2020 research and innovation programme under the Marie Skłodowska-Curie grant agreement No 861153, H2020 MeM-Scales project (871371) and  ERA-NET CHIST-ERA   programme   by   SNSF (20CH21186999 / 1).

\bibliographystyle{IEEEtran}
\bibliography{biblioncs.bib}
\end{document}